\documentclass[twocolumn,prl,reprint]{revtex4}
\usepackage{graphicx}
\usepackage{epsfig,amsmath}
\usepackage{amssymb}
\usepackage{rotate}
\usepackage{color}
\usepackage[ps2pdf=true]{hyperref}
\usepackage{bm}

\newcommand{\be}{\begin{eqnarray}}
\newcommand{\ee}{\end{eqnarray}}

\newcommand{\e}{{\mathrm{e}}}

\newcommand{\dd}{{\rm d}}
\newcommand{\vx}{{\bf x}}
\newcommand{\vk}{{\bf k}}

\newcommand{\vtau}{{\underline{\tau}}}
\newcommand{\vlambda}{{\underline{\lambda}}}
\newcommand{\vrho}{{\underline{\rho}}}
\newcommand{\vw}{{\underline{w}}}

\newcommand{\Dirac}{\delta_{\rm D}}

\newcommand{\mD}{{\cal D}}

\begin{document}
%\title{Large Deviation Principle in Large-Scale Structure Cosmology}
\title{A Large Deviation Principle at play in Large-Scale Structure cosmology}

\author{Francis Bernardeau}
\affiliation{Sorbonne Universit\'es, UPMC Univ Paris 6 et CNRS, UMR 7095, Institut d'Astrophysique de Paris, 98 bis bd Arago, 75014 Paris, France}
\affiliation{CEA - CNRS, UMR 3681, Institut de Physique Th\'eorique, F-91191 Gif-sur-Yvette, France}
\author{Paulo Reimberg}
\affiliation{Sorbonne Universit\'es, UPMC Univ Paris 6 et CNRS, UMR 7095, Institut d'Astrophysique de Paris, 98 bis bd Arago, 75014 Paris, France}
\affiliation{CEA - CNRS, UMR 3681, Institut de Physique Th\'eorique, F-91191 Gif-sur-Yvette, France}

\begin{abstract}
We present an application of Large Deviation Theory to the problem of structure growth on large-scale structure cosmology. Starting from gaussian distributed overdensities on concentric spherical shells, we show that a Large Deviation Principle holds for the densities on the corresponding shells after gravitational evolution if no shell-crossing happens. As consequences of the Large Deviation Principle we obtain the cumulant generating function for the non-linear densities, and present formulae to compute the cumulant generating function for general window functions. 
\end{abstract}


\maketitle
%%%%%%%%%%%%%%%%%%%%%%%%%%%%%%%%%%%%%%%%%%%%%%%%%%%%

The statistical properties of the large-scale structure of the universe are key to our understanding of the content of the universe, the origin of its structure, or the processes at play during the formation, evolution of the galaxies of the universe.  This is a key-stone of the science case of many on-going and future projects of cosmological observations \cite{boss, Abbott:2005bi, lsst, euclid}.

In the current scenario,  the cosmic density fields, whether related to the mass density field or the velocity field,  evolve at large-scale under the sole influence of gravity from initial seeds -- primordial metric fluctuations -- whose general properties are now well known. The initial conditions prevailing for the development of the large-scale structure of the universe  are expected to be adiabatic, identical in all fluid types, and follow, to a very good accuracy, Gaussian statistical properties. Moreover at the time non-linear couplings start to develop, it is natural to assume that only the growing mode survive. Initial conditions are then entirely determined by the field $\delta(t,\vx)$, the mass density contrast in the linear regime whose growth rate $D_{+}(t)$ is independent of scale,
\begin{equation}
\delta(t,\vx)=D_{+}(t)\ \tau(\vx)
\end{equation}
defining a field $\tau(\vx)$ (normalizing $D_{+}(t)$ to be unity at current time for definiteness). Cosmological parameters, non-standard physical effects, are expected to affect either $D_{+}(t)$ or the statistical properties of $\tau(\vx)$. The latter are better characterized by its Fourier modes, $\tau(\vk)$ and more precisely by the density power spectrum $P(k)$ defined from ensemble averages of mode products,
\begin{equation}
\mathbb{E}[\tau(\vk)\tau(\vk')]=\Dirac(\vk+\vk')P(k)
\end{equation}
where the Dirac distribution $\Dirac(\vk+\vk')$ emerges because of the statistical isotropy of the field $\tau(\vx)$. The power spectrum $P(k)$ is a complicated but smooth function of $k$ which can locally be approximated by a power law of index $n$ from $-3$ to $1$.

There are currently a wide interest in studies on how the large-scale structure of the universe evolve with time as they enter the non-linear regime, i.e.,  when the density contrast reaches values comparable to unity. One standard approach is to use a perturbation theory approach that is to compute the density field in subsequent order with respect to the initial density field $\tau(\vx)$,
\begin{equation}
\delta(t,\vx)=\sum_{n}\delta^{(n)}(t,\vx)
\end{equation}
where $\delta^{(1)}$, $\delta^{(2)}$, etc.  are respectively linear, quadratic, etc. in the field $\tau(\vx)$ (see \cite{2002PhR...367....1B, 2013arXiv1311.2724B} for details). Such expansion allows to compute next-to-leading order corrections to quantities such as the density power spectra and to compute statistical properties sensitive to mode couplings effects such as high-order correlation functions and cumulants.

Another line of investigation is the one presented in this letter. It follows early investigations by \cite{1984ApJ...279..499F,1993ApJ...412L...9J,1992ApJ...392....1B,1994A&A...291..697B,2002A&A...382..431V} where it was shown how high order correlators, in particular cumulants, could be computed from the motion equations of the fluid. The aim of this paper is to show that these results can be reformulated as the consequences of the existence of Large Deviation Principle in this system. 
%%%%
\\
\\
The Large Deviations Theory \cite{varadhan, der_hollander, ellis} deals with rate at which probabilities of certain events decay as a natural parameter of the problem varies \cite{varadhan_lec_notes}, and is applied in variety of domains in mathematics and theoretical physics, specially in statistical physics both for equilibrium and non-equilibrium systems (see for instance \cite{2009PhR...478....1T} for a review paper on the subject). For our needs on this paper, it will be sufficient to say that \emph{Large Deviation Principle} (LDP) holds for a generic family of random variables $\{\alpha_i\}$ with \emph{rate function} $I(\, \cdot \, )$ if
\begin{equation}
\label{LDP_limit}
\lim_{\epsilon \to 0} \epsilon \, \log \left[ P_{\epsilon}(\{\alpha_i\} \in A) \right]= - \inf_{\underline{\alpha} \in A} I(\underline{\alpha})
\end{equation}
where $A \subset \mathbb{R}^N$, and $\epsilon$ is the \emph{driving parameter}\footnote{To be precise, we should require two different limits to hold, one for open, and one for closed sets. The definition given, however, is consistent if we consider sets $A$ for which $\inf_{\alpha \in A^{o}} I(\alpha) = \inf_{\alpha \in A} I(\alpha) = \inf_{\alpha \in \bar{A}} I(\alpha)$, where $A^o$ is the \emph{interior} of $A$, and $\bar{A}$ its \emph{closure}. 
This condition will hold for most of the ``naturally conceivable" sets $A$, but would fail, for example, for the ternary Cantor set.}. We use here an underline to denote the vectors in $\mathbb{R}^N$, whose components will be denoted with upper indexes. Lower indexes will always label elements of families of random variables. 

Under the action of gravity linear mass density fluctuation will evolve into a non-linear field, with intricate statistical properties. We would like to obtain a LDP for a set of random variables $\{ \rho_i \}$ obtained by smoothing at some domains the non-linear field produced by the gravitational evolution of initial gaussian seeds. Establishing measures over initial configurations and assuring the existence and good behavior for solutions of the fluid equations describing gravitational evolution are, by no means, easy tasks. One way of circumvent these problems is to consider configurations that are spherically symmetric. In this case, since matter contained in a spherical cell (in a static, asymptotically flat spacetime) evolves independently of the external ambient, analytic solutions of the gravitational collapse can be obtained, and are known as \emph{spherical collapse} solutions \cite{1980lssu.book.....P}. Setting, for radii $r_1  < \ldots < r_n$, the averages:
\begin{equation}
\label{tau_i}
\tau_i := \frac{1}{\frac{4}{3} \pi r_i^3} \int_{|\vx| < r_i} \tau(\vx) \, \dd^3 \vx  \, ,
\end{equation}
the theory of spherical collapse prescribes the values of the set of non-linear densities $\{ \rho_i \}_{1 \leq i \leq N} $, at corresponding radii $R_i$, by:
\begin{equation}
\label{map}
\rho_i = \zeta(\eta, \tau_i) \qquad R_i = r_i/\rho_i^{1/3} \, .
\end{equation}
The map $\zeta$ is a cosmological dependent function \cite{1992ApJ...392....1B, 1980lssu.book.....P} well behaved at low densities, but divergent at finite values of $\tau_i$. The divergence is associated with shell crossing, and before it takes place the pairs $(\tau_i, r_i)$ and $(\rho_i, R_i)$ are in 1-1 correspondence for each $i$. 

Our goal, therefore, is to establish the LDP for the variables $\{ \rho_i \}$ obtained in spherical collapse, and to identify its rate function. This can be achieved by remarking first that the LDP holds in general for the field variables $\tau(\vx)$ and more specifically for the variables $\{ \tau_i \}$  that are all Gaussian distributed. The latter ones have the rate function $I(\vtau)$ \cite{varadhan}:
\begin{equation}
I(\vtau) = \frac{\sigma^2(r_N)}{2} \langle \vtau, \Xi \vtau \rangle \, .
\end{equation}
Here $\Xi =\Sigma^{-1}$, and $\Sigma$ it the \emph{covariance matrix} whose elements are $\Sigma_{ij} = \mathbb{E}[\tau_i \tau_j]$. The canonical scalar product in $\mathbb{R}^N$ is denoted by $\langle \, \cdot \, , \, \cdot \, \rangle$. As a consequence of the positive-definiteness of $\Sigma$, $I(\vtau)$ is a convex function of $\vtau$.

The driving parameter here is the variance associated to the random field. The joint correlators $\Sigma_{ij}$ are given by
\begin{equation}
\Sigma_{ij}=\int\frac{\dd\vk^{3}}{(2\pi)^{3}}\,P(k)\,w_{T}(k r_{i})\,w_{T}(k r_{j})
\end{equation}
where $w_{T}(k)$ is the Fourier transform of the top-hat window function\footnote{The window function is explicitly given by $w_{T}(k)=3\,j_{1}(k)/k$, with $j_1(\, \cdot \,)$ being the spherical Bessel function of order one.}, and $\sigma^2(r_i) \propto r_i^{-(n+3)}$ for a power-spectrum that scales like $k^n$. The limit $\sigma^2(r_i) \to 0$ can be performed by taking the normalization of the power-spectrum to zero, corresponding to the early stage of the dynamics or to large enough scales. Since all variances $\sigma^2(r_i)$, $1 \leq n \leq N$, behave alike on the limit $\sigma^2(r_i) \to 0$, we can take any of the variances $\sigma^2(r_i)$ as normalization factor. For definiteness, we can take the one associated to the larger radius. Because $\Xi$ scales like $\sigma^{-2}$, $I(\vtau)$ is independent of the normalization of the power spectrum. 

As the LDP holds for the variables $\{ \tau_i \}$, and the variables $\{\rho_i \}$ are image of $\{ \tau_i \}$ under the continuous map $\zeta$, we can apply a theorem on the Large Deviations Theory known as \emph{contraction principle} to affim that LDP also holds for the variables ${\rho_i }$ with rate function
\begin{equation}
\Psi(\vrho) = \inf_{\vrho = \zeta(\vtau)} I (\vtau) \, ,
\end{equation}
where $\vrho \in \mathbb{R}^N$ is a vector whose components are $\rho^i = \zeta(\tau^i)$ and each $\rho^i$ is uniquely associated to a radius $R_i$. The contraction principle essentially translates the general idea that \emph{any large deviation is done in the least unlikely of all unlikely ways}\cite{der_hollander}.

The fact that for each $i$, $1 \leq i \leq N$, $\rho^i = \zeta(\tau^i)$ irrespectively of any $\tau^j$, $j \neq i$ makes the inversion problem unidimensional: one can locally express $\tau^i = \zeta^{-1} (\rho^i)$ on the neighborhoods of point where $\frac{ d \zeta(\tau^i)}{d \tau^i} \neq 0$, what holds before shell crossing.
As a result the value of the rate function $\Psi(\vrho)$ is given by the expression of the rate function of a set of $N$ variables $\tau_{i}$ with radii $r_{i}$ prescribed by Eq. \eqref{map},
\begin{equation}
\Psi(\vrho) = \frac{\sigma^2(R_N)}{2} \langle \vtau(\vrho), \Xi \, \vtau(\vrho) \rangle \, 
\end{equation}
where the normalization factor is now chosen to be $\sigma^2(R_N)$.

A direct consequence of this setting is that we can
use Varadhan's Lemma \cite{varadhan, der_hollander}  to compute the \emph{scaled cumulant generating function} (SCGF) for the variables $\{ \rho_i \}$. If $\zeta$ is bounded, then:
\begin{eqnarray}
\label{scaled_cumulant}
\phi(\vlambda) & := & \lim_{\sigma^2(R_N) \to 0} \sigma^2(R_N) \log \mathbb{E} \left[ \e^{\sum_{i= 1}^N \lambda^i \rho_i/\sigma^2(R_N)  } \right] \nonumber\\ & = & \sup_{\vtau} \left[  \langle \vlambda , \vrho(\vtau) \rangle - \frac{\sigma^2(R_N)}{2} \langle \vtau, \Xi \, \vtau \rangle \right] \, .
\end{eqnarray}
The SCGF is, by its definition, a convex function of $\vlambda$.

While $\vtau$ can be expressed in terms of $\vrho$, we can take the supremum over the variables $\vrho$, and therefore Eq. \eqref{scaled_cumulant} takes the form of a \emph{Legendre-Fenchel} transform.
The first order condition reads
\begin{equation}
\label{first_order}
\lambda^i = \frac{\sigma^2(R_N)}{2} \frac{\partial}{\partial \rho^i} \langle \vtau(\vrho), \Xi\,  \vtau (\vrho) \rangle \, .
\end{equation}
Clearly, $\vlambda$ is uniquely determined as long as $\langle \vtau(\vrho), \Xi \, \vtau (\vrho) \rangle$ is a convex function of $\vrho$. If this is the case, Legendre-Fenchel transform reduces to a Legendre transform, and the SCGF is the Legendre transform of the rate function $\Psi(\vrho)$. If the rate function ceases of being convex, SCGF will be the Legendre-Fenchel transform of the rate function.
%%%%%%%%%%%%%%%%%%%%%
%\subsection{The case $N=1$}
%%%%%%%%%%%%%%%%%%%%%
\\
\\
In order to illustrate this construction, we shall study the problem when only one spherical shell is taken into account. This case was actually explored 
in detail in \cite{2002A&A...382..412V,  PhysRevD.90.103519}. The rate function is given by,
\begin{equation}
\label{rate_1d}
\Psi(\rho) = \frac{\sigma^2(R) \tau(\rho)^2}{2 \sigma^2(R \rho^{1/3})} \, ,
\end{equation}
where $\rho$ is associated to the radius $R$. We know that a scaled cumulant generating function $\phi(\lambda)$ will be associated to $\Psi(\rho)$. We can write its Taylor expansion as:
\begin{equation}
\phi(\lambda)=\sum_{p=1}^{\infty}S_{p}\frac{\lambda^{p}}{p!} \, ,
\end{equation}
and if $\phi$ is differentiable on a neighborhood of the origin, we can directly derive the coefficients $S_p$. We learn from convex analysis that if $\Psi(\rho)$ is convex, then $\phi(\lambda)$ is  differentiable for all $\lambda \in \mathbb{R}$. However, if $\Psi(\rho)$ fails to be convex, $\phi(\lambda)$ will have non-differentiable points. Since the coefficients $S_p$ can only be obtained if the Taylor series is defined at the origin, we have to be sure that $\phi(\lambda)$ admits derivatives at $\lambda=0$. 

\begin{figure}[!ht]
\centering
\includegraphics[width=7.5cm]{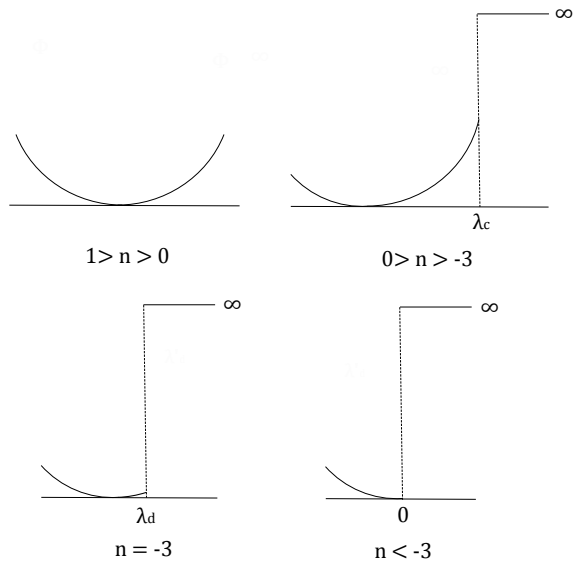}
\caption{Schematic representation of the SCGF for different ranges of $n$. For $1<n<0$ we have a convex differentiable function for all values of $\lambda$. For $0 >n>-3$ the SCGF is discontinuous at a value $\lambda_c >0$. For $n=-3$ the discontinuity appears at a value $0<\lambda_d<\lambda_c$. For $n<-3$ the SCGF is discontinuous at $\lambda=0$. In all cases the SCGF is differentiable for negative values of $\lambda$.}
\label{graphs_scgf}
\end{figure}

In order to address this point, it is convenient to remark that the function $\zeta(\tau)$ can be approximated by \cite{1992ApJ...392....1B}
\begin{equation}
\zeta(\tau) = \frac{1}{ \left( 1 -  \frac{2}{3} \tau \right)^{3/2}} \, .
\end{equation}
For $P(k) \propto k^n$, $\sigma^2(R \rho^{1/3}) \propto \rho^{-(n+3)/3}$, and the behavior of the second derivative of $\Psi(\rho)$ can be analyzed in terms of $n$.  First we remark that if $n>1$, $\Psi(0)=0$ what would lead to a bimodal PDF. We shall consider only the case $n \leq 1$, what implies that $\Psi(\rho)$ is convex for $0 \leq \rho \leq 1$. For $0 \leq n \leq 1$, $\Psi(\rho)$ is convex for all values of $\rho$. In these cases the SCGF and the rate function are Legendre conjugate of each other, and $\phi(\lambda)$ is everywhere differentiable. When $n < 0$, $\Psi(\rho)$ will no longer be globally convex, and the SCGF will be a Legendre-Fenchel transform of the rate function. If $0 > n > -3$, $\Psi(\rho)$ will be convex up to values of $\rho$ larger than $1$, but will change its concavity and grow to infinity with horizontal asymptote. If $n=-3$, $\Psi''(\rho)$ remains negative as $\rho \to \infty$, but $\Psi(\rho)$ attains a finite value at the limit. For $n<-3$, there is a second change in concavity for higher values of $\rho$, and $\Psi(\rho) \to 0$ as $\rho \to \infty$ with $\Psi''(\rho) >0$. The associated SCGF for $n < 1$ are depicted in Fig. \ref{graphs_scgf}. We see that $\phi(\lambda)$ will not admit derivatives at $\lambda = 0$ if $n < -3$. For all cases of cosmological interest, however, we have $-3 \leq n \leq 1$, and therefore $S_p$ can be obtained from the Taylor expansion of $\phi$.

In case of Gaussian initial conditions, the coefficients $S_p$  correspond to the value of the cumulants taken at leading order (i.e. at tree order in a diagrammatic point of view),
\begin{equation}
\langle\rho^{p}\rangle_{c}^{\rm tree\ order}=
\sum_{\rm configurations, \{n_{i}\}}
\left\langle \Pi_{i=1}^{p}\ \rho^{(n_{i})}
\right\rangle_{c}
\end{equation}
where the sum is performed on all configurations for which $\sum_{i}n_{i}=2p-2$, and that leads to a cumulant of the order of $\langle\delta_R^{2}\rangle_{c}^{p-1}$ \cite{1992ApJ...392....1B}. 

The direct relation between the coefficients $S_p$ and the spherical collapse problem is evident using the language of Large Deviation Theory, but requires a much more involved development to emerge on the perturbative calculation.  For instance, the density skewness of the top-hat window function is given by
\begin{equation}
S_{3}=3 \nu_{2}+\frac{\dd \log}{\dd \log R} \langle\rho^{2}\rangle_{c}
\end{equation}
and all subsequent order can then be similarly computed where the coefficients $\nu_n$ appear in the Taylor expansion of $\zeta(\tau)$, i.e., $\zeta(\tau)=\sum_{p} \nu_{p}{\tau^{p}}/{p!}$~\cite{1994ApJ...433....1B}.
%%%%%%%%%%%%%%%%%%%%%%%%%%%%%%%%%%%%%%
%\section{Density profiles}
\\
\\
Let us now turn to the density defined with a general window function defining $\rho_{R}$ and assume for the time being it can be obtained from a finite number of concentric cells of radii $R_{i}$ with the help of weights $w_{i}$,
\begin{equation}
\rho_{R}= \langle \vw , \vrho \rangle \, .
\end{equation}
As $\rho_R$ is a composition of continuous functions of $\vtau$, the contraction principle also can be applied so that $\rho_R$ will obey LDP with a rate function $\tilde{\Psi}(\rho_R) = \inf_{\rho_R= \langle \vw, \vrho \rangle} I(\vtau)$ and its scale cumulant generating function is nothing but
\begin{equation}
\phi_{s}(\lambda)=
\sup_{\vtau}\left[
\lambda \langle \vw, \zeta(\vtau) \rangle-\frac{\sigma^2_R}{2} \langle \vtau, \Xi \vtau \rangle \right] \, .
\end{equation}
Note that in the expression the inverse $\Xi$ matrix is computed for non trivial values of $r_{i}$.

It is possible however to extend the formalism to the continuous limit. The contraction principle can then be written in terms of a function $\tau(x)$. The slight difficulty we have is to express the constrain in terms of this function.
We are led to define the function 
\begin{equation}
R_{\tau}(x)=x\ \zeta^{-1/3}(\tau(x)),
\end{equation}
and the function $\xi(y,z)$ such that 
\begin{equation}
\int\dd y\ \sigma^{2}(x,y)\ \xi(y,z)=\Dirac(x-z) \, ,
\end{equation}
so that the SCGF now reads,
\begin{eqnarray}
\phi_{s}(\lambda) &=& \sup_{\tau(x)} \left[ \lambda \int\dd x\ R'_{\tau}(x)\ w(R_{\tau}(x))\ \zeta(\tau(x)) \right. \nonumber \\ && \hspace{-1.5cm} \left. -\frac{\sigma^2}{2}\int\dd x\,\dd y\,\tau(x)\,\tau(y)\,\xi(x,y) \right].
\label{phiscontinuous2}
\end{eqnarray}
When the above Legendre-Fenchel transform reduces to a Legendre transform, $\phi_{s}(\lambda)$
is obtained from a stationary condition that $\tau(x)$ should satisfy,
\begin{equation}
\tau(x)=\frac{\lambda}{\sigma^2} \int\dd y\ \sigma^2(x,y) \mD_{y}\left[ R'_{\tau}(y)\ w(R_{\tau}(y))\ \zeta(\tau(y)) \right]
\label{htauexpression}
\end{equation}
where $\mD_{y}$ is the operator,
\begin{equation}
\mD_{y} := \left( \frac{\partial }{\partial \tau}-\frac{\dd}{\dd y}\frac{\partial}{\partial \tau'}
\right).
\end{equation}

\begin{figure}[!ht]
   \centering
   \includegraphics[width=7.5cm]{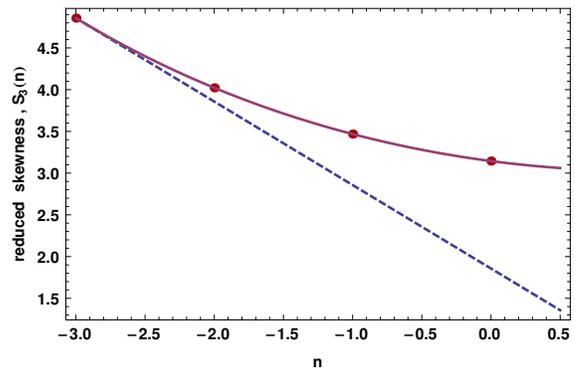}
   \caption{The skewness of the density field for Gaussian and top-hat window functions in case of power law spectra, $P(k)\sim k^{n}$. The expressions
   are given as a function of the index $n$. The dots are the results obtained in \cite{1993ApJ...412L...9J} for specific values of the index $n$.}
   \label{fig:S3}
\end{figure}

Expanding the resulting expression of $\phi_{s}(\lambda)$ close to $\lambda=0$ naturally leads to the expression of the low order cumulants. We thus have
\begin{equation}
\label{S3_gen}
S_{3}=
3\nu_{2}\frac{\int\dd x\ w(x)\ \Sigma^{2}(x)}{\left[\int \dd x\ w(x)\ \Sigma(x)\right]^{2}}+
\frac{\int\dd x\ x\ w(x)\ \Sigma(x)\ \Sigma'(x)}{\left[\int \dd x\ w(x)\ \Sigma(x)\right]^{2}}
\end{equation}
with
\begin{equation}
\Sigma(x)=\int\dd y\ \sigma^2(x,y)\,w(y)
\end{equation}
which generalizes the top-hat window result, and allows to compute low order moments for 
any type of window function. For instance, for a Gaussian window function we have,
\begin{equation}
w(x)=\frac{1}{3}\sqrt{2/\pi}\,x^{4}\,\exp\left(-\frac{x^{2}}{2}\right).
\end{equation}
from which one can compute the skewness, as shown in Fig. \ref{fig:S3}.
Note that
 Eqs. \eqref{htauexpression} and \eqref{S3_gen} are new results in the context of large-scale structure cosmology, and can be generalized for non-linear -- but bounded, mono valued and differentiable -- functions of $\{ \rho_i \}$. Besides the computation of low order cumulants, this formalism can be used to explore the analytic properties of $\phi_{s}(\lambda)$ and eventually reconstruct whole density PDFs from it. 
How can it be done? The first step is to model the cumulant generating function which can naturally be chosen
to match its $\sigma^{2}\to 0$ limit,
\begin{equation}
\phi_{\rho}(\lambda)=\frac{1}{\sigma^{2}}\phi_{s}(\lambda\sigma^{2})
\end{equation}
where here $\sigma^{2}$ can be adjusted to match the measured variance if necessary.
Then the PDF can be computed from the inverse Laplace transform of the moment generating function 
\begin{equation}
\label{PDFfromphi}
P(\rho_{R}) = \int_{-i {\infty}}^{+i \infty} \frac{\dd\lambda}{2\pi i} \exp[-\lambda\rho_{R}+\phi_{\rho}(\lambda)] \,.
\end{equation}
which in turn requires an
analytic extension of its expression in the complex plane.
Such a construction has been done extensively for the top-hat window function as in \cite{1994A&A...291..697B,1995A&A...301..309B,2000A&A...364....1B,PhysRevD.90.103519}. Our results now show that it can 
also be done for any filter shape from the numerical resolution of Eqs. (\ref{phiscontinuous2}) and (\ref{htauexpression}).
%\section{Discussion}
\\
\\
We have shown, therefore, that some key results derived for large-scale cosmic fields can be re-visited in the context of Large Deviations Theory. Whereas expressions of the density SCGFs were derived previously from direct perturbation theory calculations assuming the field density contrasts to be small everywhere, the very same results were derived here only assuming the density variance to be small. The drawback is that LDP does not offer means to implement perturbative corrections to this limit. A critical analysis of the validity of LDP results is therefore needed in general. This can be examined
at the level of the computed cumulants such as the coefficients $S_{p}$, which are measurable quantities, together with 
the cumulant generating function 
which is eventually used to reconstruct the probability distribution functions of the non-linear density field following the prescriptions described above.  Such measurements have been made in various studies \cite{1995MNRAS.274.1049B,1994ApJ...433....1B,PhysRevD.90.103519, LogDensity} and suggest that the $\sigma \to 0$ limit is robust. Such robustness is confirmed at the level of  the reconstructed PDFs for filtered densities at small but finite variances as presented in \cite{PhysRevD.90.103519, LogDensity}.

Note that in the case of top-hat filtering, the SCGF can be extended everywhere in the complex plane (not being analytic along the real axis) after regularization of the spherical collapse. In previous works \cite{PhysRevD.90.103519}, this analytical continuation was used to get the PDF for collapsed densities.

One way of avoiding the complications introduced by discontinuity of the SCGF is to perform ``changes of variables" that push the non-convexity of the rate function to high values of the density. Such a construction is explored in \cite{LogDensity}, where the PDFs for the logarithm of the density are reconstructed, and contrasted with the ones obtained directly from the density. 

Besides novel insights into the nature of the results that have been found, this formalism gives a better idea of the reach of applicability of such developments.  It can, for instance, be directly extended to the velocity field divergence \cite{1992ApJ...390L..61B}, or could even be applied after shell crossing provided the spherical collapse dynamics is regularized (though the determination of the least unlikely configurations for large deviation to happen might then be difficult). Among the other domains of applicability one could mention the case of primordial non-gaussianities, field configurations with other types of symmetries such as cylindrical symmetry (as in \cite{1995A&A...301..309B,2000A&A...364....1B}). One could also apply the formalism to derive statistical properties of wide range of observables provided they can be expressed, not necessarily linearly, in terms of the densities in concentric cells.
\\
%\subsection*{Acknowledgements} 
This work was supported by the grant ANR-12-BS05-0002 and Labex ILP (reference ANR-10-LABX-63) part of the Idex SUPER of the programme Investissements d'avenir under the reference ANR-11-IDEX-0004-02.
\bibliographystyle{OUPnamed}
\bibliography{LSStructure}

%\appendix
 
 \end{document}